\documentstyle[aps,prl]{revtex}
\title{Emission of light mesons directly from the  surface of quark-gluon
        plasma.}
\author{D.Yu. Peressounko and Yu.E. Pokrovsky}
\address{Russian Research Center "Kurchatov Institute"}
\begin{document}
\maketitle

\begin{abstract}
On the basis of hydrodynamic model of evolution we consider emission of
lightest ($\pi,K,\eta,\rho,\omega,K^*$) mesons directly from the surface of
quark-gluon plasma, created in the heavy ion collision, with accounting of
their absorption by surrounding hadronic gas.  We evaluate upper and lower
limits on yields of these direct mesons in Pb+Pb collisions at SpS, RHIC and
LHC energies, and find, that even in the case of the lowest yield, direct
$K$, $\eta$ and heavier mesons dominate over freeze-out ones at soft $p_t$
($p_t\le 0.5\,GeV/c$). This leads to enhancement of the low $p_t$ production
of these mesons which hardly can be explained within pure hadronic gas
scenario, and can be considered as quark-gluon plasma signature.

\end{abstract}

\pacs1{25.75.-q,12.38.Mh,25.75.Dw,24.10.Nz}

As a rule, considering heavy ion collision with QGP creation, one assumes,
that hadrons, produced on plasma surface, suffer numerous rescatterings in
the surrounding hadronic gas, so that final hadrons do not carry direct
information about plasma. In this letter we show, that this widely accepted
opinion is not quite right for the case of finite systems, created in heavy
ion collisions. Because of comparability of free path lengths of hadrons in
the hadronic gas with sizes of the region, occupied by hot matter,
significant number of hadrons, created on the surface of quark-gluon plasma
can pass through the surrounding hadronic gas without rescattering. Such kind
of final state hadrons we call {\it direct} hadrons below. In contrast to the
usually considered {\it freeze-out} hadrons, originated due to evaporation
from the hadronic gas or its freeze-out, direct hadrons carry immediate
information about the plasma surface. The goal of this letter is to show,
that the yield of direct mesons is not negligible small, and moreover, there
is kinematic region, where direct hadrons dominate over freeze-out ones.

The rather obvious itself fact, that final hadrons are emitted not from thin
freeze-out hypersurface, but from the whole volume, occupied by hadronic gas,
was noted several times within different models of AA collisions: within
quark-gluon string model \cite{Bravina}, within the model, combining
hydrodynamic description of evolution of QGP and relativistic quantum
molecular dynamics for hadronic gas \cite{Hydro+RQMD}, within similar to our
approach, used in papers \cite{Grassi} for evaluation of hadronic emission
from the depth of hadronic gas. Nevertheless, until our paper \cite{We},
there were no attempts to separate direct hadrons from QGP surface from ones
from hadronic gas.  Surely, one can not point out hadron and tell, that it
e.g. came from QGP surface, but as we will show, it is possible to find
kinematic region, where hadrons from QGP surface dominate. In the paper
\cite{We} we considered S+Au collision at $200\,A\cdot GeV$ (SpS) and
evaluated yields of direct and freeze-out pions. We showed that direct pions
can dominate in soft $p_t$ region, what result in enhancement of the yield of
pions with low $p_t$.  However, there are a lot of effects, such as resonance
decay, absence of chemical equilibrium etc., which lead to the similar
enhancement of pion spectrum, making strong physical background for direct
pions. In this letter, in addition to pions, we evaluate emission of heavier
direct and freeze-out mesons:  $K,\eta,\rho,\omega$ and $K^*$ in AA
collisions at SpS, RHIC and LHC energies, and demonstrate, that for heavier
mesons contribution of direct mesons is even larger than for pions while
physical background is negligible.

To estimate yields of the direct mesons in heavy ion collision we use the
following model. Hot matter, created in the very beginning of collision,
evolves hydrodynamically. On the background of this evolution the direct
mesons are continuously emitted from the surface of QGP as a result of flying
out of quarks and gluons from the depth of quark-gluon plasma, their
hadronization on the plasma surface and fly out of the direct mesons through
surrounding expanding hadronic gas sometimes with rescattering. If direct
meson suffers rescattering in the hadronic gas, then we assume, that it lost
direct information about plasma and consider it further hydrodynamically. For
freeze-out hadrons we assume thermodynamic and chemical equilibrium at
freeze-out moment. More elaborated description of freeze-out of hadronic gas,
such as e.g. \cite{hadron-freeze-out}, can only increase the relative yield of
direct hadrons.

Probability for quark and gluon being emitted in the depth of QGP to reach
its surface, and for meson to escape from hadronic gas without rescattering
is determined by expression:
$$
P=\exp \left\{ -\int \lambda ^{-1}(\varepsilon ,x)\,dx\right\},
$$
where integration is performed along the path of the particle in the hot
matter with accounting of its evolution, and $\lambda (\varepsilon ,x)$ -
free path length of the particle, which depends on energy of the particle
$\varepsilon$ and local energy density at the point $x$. We calculate free
path lengths of quark and gluon in QGP and meson in hadronic gas using
equation

$$
\lambda _{i}(\varepsilon )=\left[ \frac 1{16\,\pi
^3}\frac T{\varepsilon \ p}\sum_j\int\limits_{(m_1+m_2)^2}^\infty \sqrt{
s^2-2\,s\,(m_1^2+m_2^2)+(m_1^2-m_2^2)^2}\,
\sigma _{ij}(s)\;\ln  \left( \frac{1-\exp (-a_{+})}{1-\exp
(-a_{-})}\right) \,ds\right] ^{-1},
$$
$$
a_{\pm }=\frac{\varepsilon(s-m_1^2-m_2^2) \pm \sqrt{(\varepsilon
^2-m_1^2)( s^2-2\,s\,(m_1^2+m_2^2)+(m_1^2-m_2^2)^2)}
}{2\,m_1^2\,T},\quad
$$
where $\sigma _{ij}(s)$ -- total cross-section of interaction of $i$ and $j$
particles, $m_1$ and $m_2$ -- mass of projectile and target particles
correspondingly, $T$ -- temperature, and sum is taken over all possible
two-particle reactions. Evaluating free path lengths of hadrons in hadronic
gas we take into account only rescatterings on pions. In the case of the free
path length of pions we use experimental cross-sections of $\pi\pi$
scattering (see \cite{We}), while for $K,\,\eta, ...$ we assume
$\sigma=\sigma_{\pi^+\pi^+}\sim 10\, mb$ plus contributions from
excitation of resonances. One could expect, that presence of nucleons in the
hadronic gas results in significant reduction of the free path length of the
pion with respect to pure pionic gas (due to excitation of $\Delta$
resonances). However, below we concentrate on the midrapidity region, where
net baryon density is small, and for reasonable values of barionic
chemical potential $\mu\le 300\,MeV$ and temperatures below $\sim 200\,MeV$,
we find, that contribution of nucleons into free path length of pions is
negligible.

Emission rate (the number of particles, emitted from unit volume per unit
time) of quarks and gluons from QGP we find from the condition, that
infinitely thick layer of QGP emits quarks and gluons in accordance with
Stephan-Boltsman formula.  So we find:

$$
\varepsilon \frac{d^7R_i}{d^3p\,d^4x}=\frac{d_i}{\lambda
_i(\varepsilon )}\frac \varepsilon {(\exp (\varepsilon /T)\pm 1)}.
$$
where $d_i$ -- degeneracy, $\lambda _i$ -- free path length of the particle,
$T$ -- temperature.

In this letter we interest in $p_t$ distributions at midrapidity region,
therefore we restrict ourselves by Bjorken 2+1 hydrodynamics with transverse
expansion, which provides good description of evolution of hot matter at this
region.

As far as it is not possible to describe consistently hadronization of quarks
and gluons on the plasma surface, we estimate upper and lower limits on the
yields of direct hadrons, by use of two models of hadronization -- `creation'
model, giving the lower limit and `pull in' model, used in the paper
\cite{We} and giving the upper limit. In both models we assume, that a quark
or gluon, flown through the plasma surface, pulls tube (string) of color
field. Creation of the final hadron corresponds to the breaking of the tube
due to discoloring of the moving out quark or gluon. In the first model we
assume, that this discoloring takes place as a result of creation of a
quark-antiquark pair in the strong field of the tube. This assumption is used
in the well known Lund model, and implemented in the event generator JETSET
\cite{Jetset}, where all parameters are chosen to fit the $e^+e^-$
annihilation at $\sqrt{s}=30\,GeV$.  Therefore, in the `creation' model we
extract corresponding probabilities from JETSET 7.4. In the second model we
assume, that discoloring takes place as a result of `pulling in' of soft
quark or gluon with corresponding color from the pre-surface layer of QGP
into the tube. Because of the large number of soft quarks and gluons in the
QGP {\it each} moving out quark and gluon can transform into some hadron,
providing its energy is larger than mass of this hadron.  If several hadrons
can be formed, then we take their relative yields from $e^+e^-$ annihilation
at $\sqrt{s}=30\,GeV$. The main difference between these two models is in the
probability of fragmentation of soft quarks and gluons: in the `pull in'
model probability is independent on energy, while in the `creation' model
probability of creation of quark pair in strong field is proportional to $\exp(
- \varepsilon ^2/p_0^2)$, where $\varepsilon$ - energy of the quark and
$p_0\sim 0.5 \,GeV$. Having probabilities of hadronization of quarks and
gluons into hadron $R_h^{q,g}(\varepsilon_q)$, evaluated using these two
models, we obtain the following hadronization function:

$$
f_h^{q,g}(\varepsilon_h,\theta_h,\phi_h)=2\, R_h^{q,g}(\varepsilon_q)
\frac{\delta(\phi_h-\phi_q)\,\delta(\cos\theta_h-\cos\theta _q)\,
      \theta(\varepsilon_q-m_h)}
     {\varepsilon_q\sqrt{\varepsilon_q^2-m_h^2}-m_h^2\ln\left
      (\varepsilon_q/m_h+\left .\sqrt{\varepsilon_q^2-m_h^2}
      \right/m_h\right ) }
$$
where $\varepsilon,\theta,\phi$ energy, polar angle and azimuthal angle of
initial quark ($q$) or gluon ($g$) or final hadron ($h$) correspondingly, and
$m_h$ -- mass of the hadron. This fragmentation function is normalized to
describe fragmentation of one quark or gluon to $R_h^{q,g}(\varepsilon_q)$
hadrons with energy in the range $m_h<\varepsilon_h<\varepsilon_q$.

To apply our model to central $Pb+Pb$ collisions at $158\, A\cdot GeV$ we
choose initial conditions (initial temperature $T_{in}$, time of
thermalization $\tau_{in}$ and initial radius $R_{in}$), QGP-hadronic gas
transition temperature ($T_c$) and freeze-out temperature ($T_f$) to
reproduce experimental $p_t$ distribution of $\pi^0$ at midrapidity
\cite{SpS-pi0-exp}. So we use:
$$
T_{in}=350\,MeV, \, \tau_{in}=0.25\,fm/c, \,
R_{in}=6.5\,fm,\, T_c=160\, MeV,\,T_f=140\, MeV.
$$

Evaluated $p_t$ distributions of various mesons at midrapidity for the `pull
in' and `creation' models of hadronization of quarks and gluons are shown on
the fig.\ 1. We do not distinguish isotopic projections of mesons, so we show
distributions, averaged over all pions, kaons etc.

\begin{figure}
\vspace{11. cm}
\includegraphics{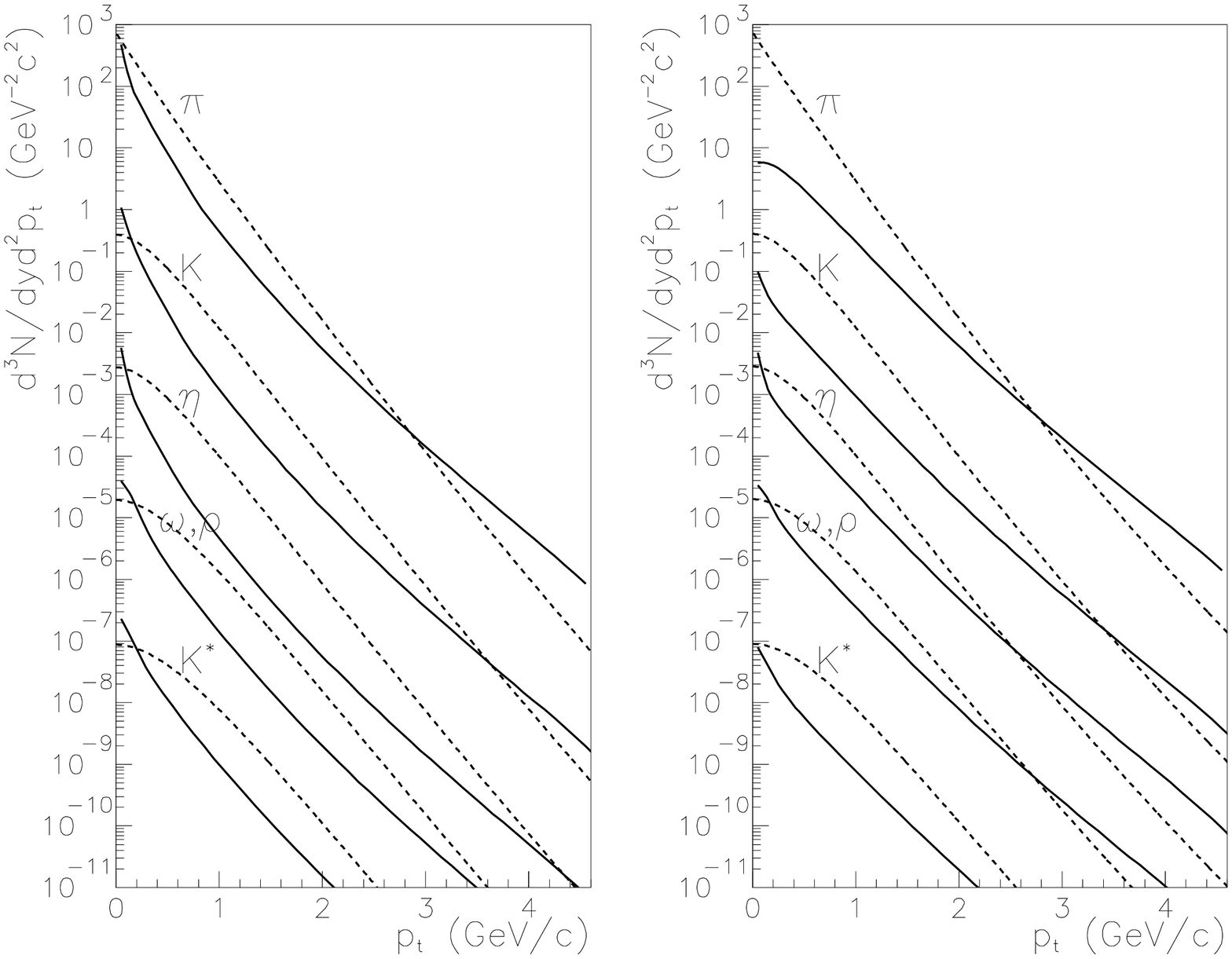}
\caption{Yields of direct (solid lines) and freeze-out (dotted lines) mesons
in Pb+Pb collision at SpS energy within `pull in' (left plot) and `creation'
(right plot) models of hadronization of quarks and gluons on plasma surface.
Distributions of $K,\eta,...$ are multiplied by $10^{-2},10^{-4},...$.}
\end{figure}

We find, that the yield of direct mesons is comparable with the yield of
freeze-out ones, see fig.\ 1. The contribution of direct mesons is higher
within `pull in' model, what is the consequence of higher probability of
fragmentation of low energy quarks and gluons in the this model of
hadronization. Nevertheless, within both models of hadronization direct
$\eta$ and heavier mesons dominate over freeze-out ones at soft $p_t$. Thus
we obtain unexpected result, that direct mesons, emitted from the hottest
region of the collision, dominate not only in the hard part of spectrum, but
also in the soft $p_t$ region. This takes place because quark or gluon,
hadronizing on the plasma surface, spreads its energy between the direct
hadron and a part of color tube, which is pulled back to plasma, what leads
to non-thermal spectrum of direct hadrons.

To investigate dependence of the yields of direct mesons on the energy of
collision, we evaluated these yields in the central $Pb+Pb$ collisions at
$3100+3100\, A\cdot Gev$ (LHC). To do this we used the following initial
conditions:
$$
T_{in}=1\,GeV, \, \tau_{in}=0.15\,fm/c, \,
R_{in}=6.5\,fm.
$$
what corresponds to the multiplicity at midrapidity $dN/dy\sim  10^4$,
predicted by cascade models, while transition and freeze-out temperatures
remains the same as for SpS. Distributions of direct and freeze-out hadrons,
evaluated in this case are shown on fig.\ 2.

\begin{figure}
\vspace*{11. cm}
\includegraphics{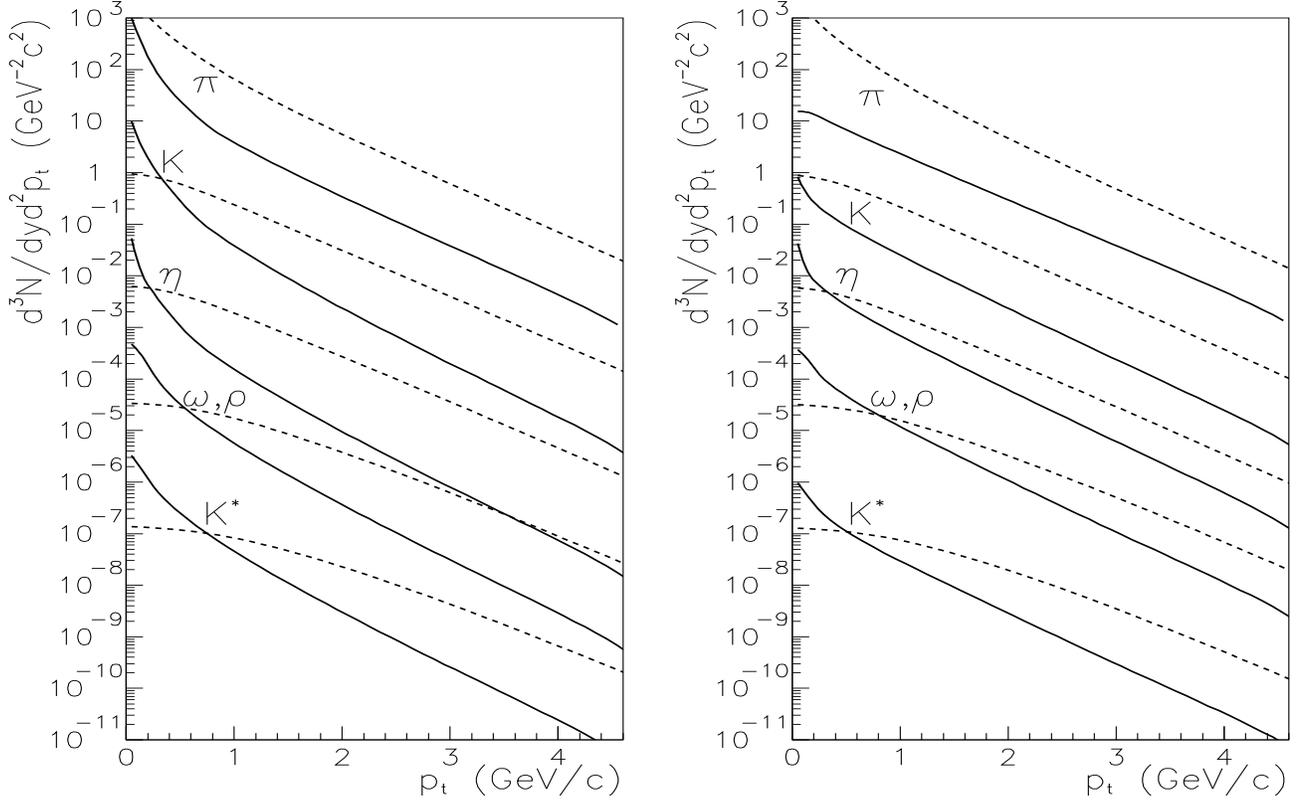}
\caption{Yields of direct (solid lines) and freeze-out (dotted lines) mesons
in Pb+Pb collision at LHC energy within `pull in' (left plot) and `creation'
(right plot) models of hadronization of quarks and gluons on plasma surface.
Distributions of $K,\eta,...$ are multiplied by $10^{-2},10^{-4},...$.}
\end{figure}

In contrast to SpS, direct hadrons do not dominate at hard $p_t$ at LHC
energy, because at this energy collective radial velocity of hadronic gas
compensate slightly higher temperature in the pre-surface layer of QGP from
which direct hadrons are emitted. In return, due to longer QGP phase
direct mesons more significantly contribute to low $p_t$ region. Again, for
$\eta$ and heavier mesons direct hadrons dominate at soft $p_t$ within both
models of hadronization.

For the sake of brevity we do not present here plots for RHIC energy, but
the contribution of direct hadrons  can be easily estimated using figs.\ 1
and 2 and Table 1.

We summarized the parts of direct mesons in the total meson yield, evaluated
for two models of hadronization for SpS, RHIC and LHC energies on the Table
1. As one can see, the part of direct hadrons is considerable and increase
with increasing of the energy of collision.

\begin{table}
\caption{Part of direct mesons in the total meson yield in midrapidity
region.}

\begin{tabular}{|c|c|c|c|c|c|c|}
\hline
    &\multicolumn{2}{|c|}{SpS} &\multicolumn{2}{|c|}{RHIC}&\multicolumn{2}{|c|}{LHC}  \\
\cline{2-7}
meson   &`pull in'&`creation'&`pull in'&`creation'&`pull in'&`creation' \\
\hline
$\pi^0$ & 0.31    & 0.04     & 0.27    & 0.04     &  0.12   &   0.03    \\
$K$     & 0.38    & 0.10     & 0.4     & 0.11     &  0.35   &   0.13    \\
$\eta$  & 0.25    & 0.31     & 0.27    & 0.35     &  0.21   &   0.37    \\
$\rho,\, \omega$
        & 0.33    & 0.36     & 0.39    &  0.45    &  0.42   &   0.52    \\
$K^*$   & 0.35    & 0.19     & 0.43    &  0.26    &  0.44   &   0.34    \\
\hline
\end{tabular}
\end{table}

We find, that direct hadrons considerably contribute to the total hadronic
yield: contributions of direct $\eta$ and heavier mesons reach $\sim 0.5$ of
the total yield. Direct mesons contribute not only to the hard part of
spectrum (at SpS energies) but also at low $p_t$.  We would like to stress,
that domination of direct hadrons in the soft $p_t$ region is not artifact of
our model, but the consequence of two rather general physical requirements.
First -- a part of the energy of quark or gluon is lost during their
hadronization.  Second -- there is approximate thermal equilibrium, to that
extend, which is usually observed in heavy ion collisions. As for the value
of this domination, it certainly depends on the details of the model: on the
probability of hadronization, transparency of hadronic gas and details of
evolution of the hot matter. We estimated sensitivity of the yield of direct
hadrons to these parameters. In this letter we evaluated upper and lower
limits for the probability of hadronization, while in the paper \cite{We} we
estimated sensitivity to variations of free path lengths and hydrodynamic
parameters. As a result, we find that the yield of direct hadrons is rather
stable with respect to variation of model parameters, and can not be changed
considerably within reasonable models.

Contribution of direct hadrons change the shape of the meson spectrum, what
can be used, in principle, as quark-gluon plasma signature. But to do this one
have to answer the question: `is it possible to find such enhancement within
pure hadronic picture of the collision?'. Fortunately for us, excess of low
$p_t$ pions in AA collisions was observed experimentally long ago
(see e.g. \cite{pi-low-exp}) and caused active discussion of possible sources
of such excess in literature. The review of proposed reasons for such excess
can be found in \cite{review}. It includes:
\begin{itemize}
\item Resonance decays \cite{pi-low-res}, mainly $\Delta$ and $N^*$. This
effect is very important for pions, especially at target and projectile
regions, however, it brings negligibly small contribution to spectra of
$\eta$ and heavier mesons. The number of heavier resonances, which can decay
onto  mesons we consider, multiplied by branching of this decay is well below
the number of direct mesons (see Table 1).

\item Collective motion, as was demonstrated in \cite{pi-low-coll} explained
observed enhancement, but later it was shown \cite{pi-low-coll-2}, that this
explanation was consequence of rather specific and unnatural freeze-out
conditions, used in \cite{pi-low-coll}. Within more physical hydrodynamic
models radial collective expansion does not bring such contribution.

\item Absence of chemical equilibrium in pion gas \cite{pi-low-chem}. This
effect essentially uses suppression of channels, changing the number of pions
with respect to elastic scattering. In the case of heavier mesons this is not
the case, so this effect does not contribute as well.

\end{itemize}

Therefore we find that excess of low $p_t$ $\eta$ and heavier mesons can not
be explained within pure hadronic scenario of collision, and, possibly, can be
considered as quark-gluon plasma signature.

To conclude, we consider emission of light ($\pi\, K,\, \eta,\, \rho,\,
\omega,\, K^*$) mesons directly from the surface of quark-gluon plasma in the
case of its creation in nucleus-nucleus collision. These mesons are
continuously emitted from  the surface of QGP due to fly-out of quarks and
gluons from pre-surface region, their hadronization and subsequent escape
from hot matter without interactions. Direct mesons give unique opportunity
to test the pre-surface layer of QGP via strongly interacting particles. We
evaluated yields of direct and freeze-out mesons in SpS, RHIC and LHC
energies, and find, that direct mesons considerably contribute to the total
meson emission -- their contributions reach $\sim 0.5$ of total yield.
Moreover, direct hadrons dominate over freeze-out ones in the soft part of
spectrum, what result in enhancement of low $p_t$ $\eta$ and heavier meson
productions, what, as we have shown, can not be explained within pure hadronic
gas scenario and can be considered as QGP signature. We argue, that such effect
is the result of two natural assumptions: first -- there is approximate
thermodynamic equilibrium, that is, there is no drops of matter with
extremely low temperature; second -- quark or gluon loses part of its energy
during hadronization on the plasma surface.  The value of enhancement depends
on the details of the model, however, one can not change it significantly
within reasonable models.

This research was supported by grants RFFI 96-15-96548 and INTAS 97-158.

\end{document}